%% file: yamato_e.tex
\documentclass{article}

\usepackage{amsmath}
\usepackage{amssymb}
\usepackage{graphicx}
\allowdisplaybreaks
\usepackage{epsf}

\usepackage[dvips]{color}

\begin{document}

\begin{titlepage}

\title{
\vspace{1cm}
{\bf
The Laue pattern and the Rydberg formula\\ in classical soliton models
}}

\author{
 Shinichiro {\sc Yamato}\,\thanks{
{\tt yamato4389@gmail.com}}
\\[7mm]
}

\date{{\normalsize January 2009}}
\maketitle

\begin{abstract}
\normalsize
In recent researches of the dynamics of solitons,
it is gradually revealed that
oscillation modes play a crucial role
when we analyze the dynamics of solitons.
Some dynamical properties of solitons on external potentials
are studied with both numerical methods and analytical methods.

In this paper, we propose a method to deal with
such oscillation modes of solitons in potential wells.
We show that oscillations of a soliton is described by
the Klein-Gordon equation with an external potential.
Although this analysis does not seems to give
quantitative scattering amplitude of a soliton itself,
it explains qualitative pictures of scattering.
As a result of our analysis,
when a soliton is scattered in a cyclic potential,
the Laue pattern emerges.
Furthermore, since our analysis is based on the Klein-Gordon equation,
a discrete frequency spectrum of a soliton is obtained
when it is bounded by some potentials.
What is especially important is that this analysis predicts
a frequency spectrum of a soliton in the Coulomb potential
and then we find that this system absorbs
external waves with specific frequencies
described by the Rydberg formula.
\end{abstract}

\thispagestyle{empty}
\end{titlepage}


\section{Introduction}
 \label{sec:intro}

Since J.~C.~Maxwell proposed the electromagnetic field theory,
many people tried to describe particles as solitons of
force fields,
but all challenges failed: They could not even construct
a solitary wave in 3+1 dimensions.

However, in 1962, the theory of hadrons proposed by T.~H.~R.~Skyrme\cite{Skyrme},
in which baryons are described as solitons of meson fields,
attracted much interest.
In this theory, he succeeded to construct a stable soliton solution,
and by the later studies it was cleared that this model gives good predictions
to the properties of hadrons\cite{ANW}.

In these days we know many soliton models in various dimensions of space-time,
but they are not neither accepted as fundamental physical theories nor denied.
That is because we don't know properties of such models enough.
By recent researches,
we established various methods to obtain static properties of solitons,
but still we do not know how we can analyze dynamical properties of them,
such as cross sections,
since no systematic method is established
to deal with time evolution of soliton models,
which are generally constructed of highly non-linear equations of motion.

Although we do not have standard methods to calculate dynamics of solitons
other than numerical calculation,
there are several reports\cite{Kevrekidis, PietteWard, PietteZakrzewski}
which show the importance of vibration modes of solitons
when we analyze the scattering of solitons.

In this paper, we propose a method to deal with
oscillation modes of dynamical solitons.
Although detailed complex structure of dynamics of solitons
is not clear in our analysis,
we believe that the qualitative pictures of dynamical solitons
are well described.

In the later sections,
we first derive an equation of an oscillation mode,
and then we argue how the equation is modified under potential wells.
With some approximation,
we see that our equation is the same as Klein-Gordon equation with potential.
Then we analyze this equation in two types of potentials,
a cyclic potential and the Coulomb potential.
Of course we have already know the results in these potentials,
but we add some comments to them.
In the section \ref{sec:summary},
we summarize our results and discuss problems of our analysis.


\section{Oscillation of a soliton}
 \label{sec:oscillation}

In this paper, we do not restrict ourselves to specific soliton models,
but we treat general properties of any soliton models.

To begin with, we assume some conditions.
First, we assume that a soliton solution in our interest has discrete
oscillation modes, especially the lowest one with frequency $\omega _0$.
\begin{figure}[h]
\begin{center}
 \scalebox{1.0}{
  \input{oscillation.pstex_t}
 }
 \caption{an image of oscillation}
 \label{fig:oscillation}
\end{center}
\end{figure}
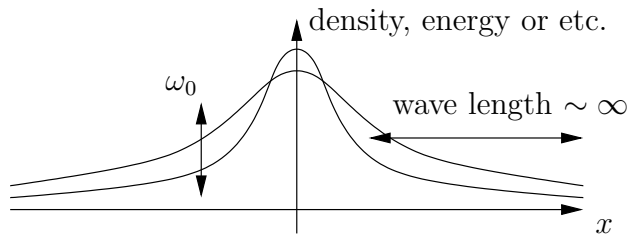

As seen from figure \ref{fig:oscillation},
usually wave length of the lowest mode is infinity.
Soliton solutions may have other oscillation modes,
but restricting ourselves to situations
in which we can neglect such high frequency modes,
we consider only $\omega _0$ in the following discussions.

The second condition is about time scale of interactions.
Generically, oscillations around a classical soliton solution
decay with specific lifetime.
In the following discussions, we assume that time scale of interactions between
solitons and something is shorter than the lifetime of the oscillation mode,
then we neglect effects of such decay.

\begin{figure}[h]
\begin{center}
 \scalebox{1.0}{
  \input{boost.pstex_t}
 }
 \caption{Lorentz boost of oscillation}  
 \label{fig:boost}
\end{center}
\end{figure}
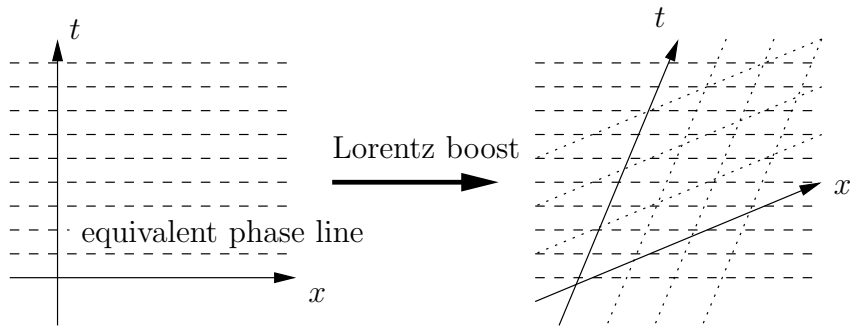

Now we begin to analyze this oscillation mode $\omega _0$.
Though amplitude of oscillation depends on space coordinates,
we approximate roughly and neglect the dependence.
Then we write the oscillation of a soliton as
\begin{align}
\phi (x, t) \cong \sin( \omega _0 t) . \label{eq:sin}
\end{align}
Since we need a moving soliton to study its dynamics,
we boost this oscillation and obtain a following wave.
\begin{align}
\phi (x, t) = \sin( \omega _0 \gamma t - \omega _0 \beta \gamma x) . \label{eq:boosted_sin}
\end{align}
This equation means that boosting a wave having infinite wave length,
we obtain a wave having a finite wave number $k = \omega _0 \beta \gamma$.
We can understand this fact intuitively by considering figure \ref{fig:boost}.
We note that the wave number $k$ is proportional to the momentum of
the soliton itself.
Actually, when we write the energy (rest mass) of the soliton as $m$,
its momentum is written as $p = m\beta \gamma$.
So we define the proportionality constant as following.
\begin{align}
k &= \omega _0 \beta \gamma \nonumber \\
  &= \frac{1}{\hbar} p \label{eq:hbar},
\end{align}
where $\hbar$ is not the Plank constant
but a model-specific constant.\footnote{For example, it is a free parameter in Skyrme model and we can make it as the real Plank constant.}
Then $m = \hbar \omega _0$ and
thus we obtain relations similar to that of quantum theory.

We can easily write an equation of this oscillation $\phi$ as
\begin{align}
\left( \partial ^2 + \omega _0 ^2 \right) \phi = 0 . \label{eq:wave_eq} 
\end{align}
Then we conclude that the oscillation modes of the soliton solution
satisfy the free Klein-Gordon equation when nothing disturbs motion of the soliton.


\section{Soliton in potentials}
 \label{sec:potential}

Next we want to derive an equation of the oscillation mode
which is also valid in potential wells.

We consider situation that a soliton pass through some potentials.
For example, we introduce potential energy to an action like
\begin{align}
S = \int d^4 x\,\mathcal{L}_0 + \int d^4 x\,J^0(x) V(x), \label{eq:interaction}
\end{align}
where $\mathcal{L}_0$ is the original Lagrangian of the soliton model,
$J^0(x)$ is the density of the soliton
and $V(x)$ represents a potential well.

\begin{figure}[!h]
\begin{center}
 \scalebox{0.7}{
  \input{potential.pstex_t}
 }
 \caption{soliton passing through step function potential}
 \label{fig:step}
\end{center}
\end{figure}
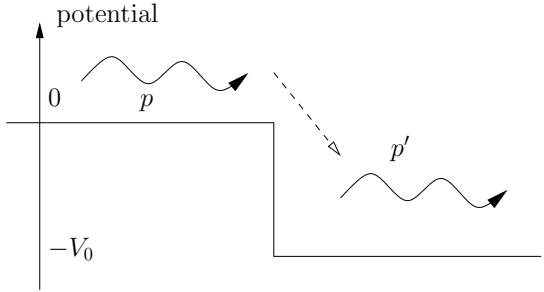

First we analyze motions of the soliton
in the case that a potential has shape of a step function
as shown in figure \ref{fig:step}.
In figure \ref{fig:step}, the oscillating soliton comes with momentum $p$,
falls the potential of depth $-V_0$ and finally gets momentum $p^{\prime}$.
By considering energy conservation, we obtain a following relation.
\begin{align}
\sqrt{m^2 + p^{\prime 2}} - V_0 = \sqrt{m^2 + p^2}. \label{eq:energy_conservation}
\end{align}
Of course exactly speaking, we must take energy of
oscillation and reflection wave into account,
but here we approximate roughly again.

Now we can modify the equation \eqref{eq:wave_eq} to satisfy the relation \eqref{eq:energy_conservation} as
\begin{align}
\left\{ \left( i\hbar \frac{\partial}{\partial t} + V(x) \right) ^2 - \left( \frac{\hbar}{i} \nabla \right) ^2 - m^2 \right\} \phi = 0, \label{eq:KleinGordon}
\end{align}
where we adjusted notations to make it clear that
this equation is the same as the Klein-Gordon equation with potential wells.
Then we can use well-known results of this equation to analyze
oscillation of solitons.


\subsection{In a cyclic potential}

\begin{figure}[ht]
\begin{center}
 \scalebox{1.0}{
  \input{cyclic.pstex_t}
 }
 \caption{soliton scattering in a cyclic potential
          (These circles are contors of the cyclic potential.)}
 \label{fig:cyclic}
\end{center}
\end{figure}
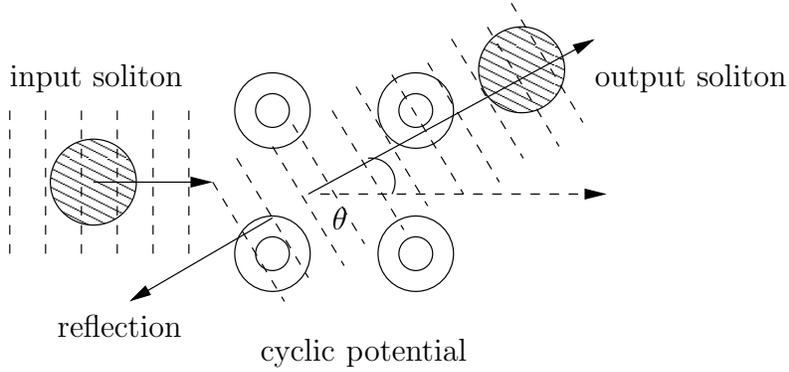

First we consider an oscillating soliton in a cyclic potential shown in figure \ref{fig:cyclic}.
By calculating transmission and reflection rates
in the equation \eqref{eq:KleinGordon},
we can easily show that the oscillation mode makes the Laue pattern.
Although what just we want to know is flux of a soliton itself,
the equation \eqref{eq:KleinGordon} seems to tell us
no information on motion of the soliton.
In this paper we try to obtain qualitative pictures of soliton flux
instead of quantitative one.

Solitons are scattered and change their momentum
when they interact with the cyclic potential.
Since we can not calculate detailed scattering phenomena of solitons
with the equation \eqref{eq:KleinGordon},
we assume that solitons are scattered to all directions
with almost uniform possibility.
When we choose one direction,
it determine a period of the cyclic potential,
and then transmission and reflection rates of oscillation
depending on the wave length and period of potential
are calculated in a standard manner.
Then we see that there exists directions with
a high transmission rate and a low transmission rate.
If a soliton is scattered to the path of a low transmission rate,
it lose its oscillation energy and kinetic energy
and eventually the flux of the soliton to that direction decreases.
As a consequence of this scattering with the cyclic potential,
we obtain the Laue pattern.

Although our analysis may not be so accurate,
actually, the fact that the length of potentials
makes a pattern of output soliton momentum in a numerical calculation
is reported in \cite{PietteZakrzewski}.
So the Laue pattern will emerge if we calculate this system numerically.


\subsection{In the Coulomb potential}

Next we consider a system in which a soliton is bounded
in the Coulomb potential.
It is well known that the quasi-static solutions of
the equation \eqref{eq:KleinGordon} with the Coulomb potential
yield eigenvalues that express the spectrum of the hydrogen atom.

When we write the Coulomb potential as $ V(x) = - \frac{\alpha}{r}$,
then the eigenvalues of the equation \eqref{eq:KleinGordon} are
given as
\begin{align}
\omega ( n_r, l )
 &= \pm \frac{m}{\hbar} \left[ 1+ \frac{\alpha ^2}{\hbar ^2 (1 + n_r + \nu )^2} \right] ^{-\frac{1}{2}} \label{eq:hydrogen} \\ 
 & \cong \pm \frac{m}{\hbar} \left[ 1 - \frac{1}{2} \frac{\alpha ^2}{\hbar ^2 ( 1 + n_r + \nu) ^2} \right] , \label{eq:hydrogen_nr}
\end{align}
where
\begin{align}
\nu = \sqrt{\left( l + \frac{1}{2} \right)^2 - \frac{\alpha ^2}{\hbar ^2}} - \frac{1}{2}. \label{eq:angularmomemtum}
\end{align}
Since we do not quantize this system,
this spectrum is not interpreted as a energy spectrum
but just a spectrum of frequency of soliton oscillation,
so we do not concern the sign of eigenvalues.

Now we conclude that this system absorbs external waves
which have frequency of following forms
\begin{align}
\omega = R \left( \frac{1}{N^2} - \frac{1}{M^2} \right) .
\end{align}
Thus, we obtain the Rydberg formula
and find that the system, a soliton in the Coulomb potential
explains the spectrum of the hydrogen atom except for effects
with which spin is concerned.


\section{Summary and discussions}
 \label{sec:summary}

In the recent studies of soliton models,
it is gradually cleared that scattering properties of solitons
have a deep connection to their oscillation modes.
In the arguments above,
we cleared that the oscillation modes of a soliton in
any classical soliton models
are described by the Klein-Gordon equation under some assumptions.
Then we researched on the possibility whether
classical soliton models can predict phenomena
which are calculated by the Schr\"odinger equation.
If a classical soliton model appear to be equivalent
to the Schr\"odinger equation,
then we are just released from the fatal problem of the quantum theory,
the measurement problem.
Furthermore, soliton models have possibility to solve many problems
on the standard model,
such as the number of parameters, the structure of generations and etc.

For this purpose, we analyzed oscillation of solitons
in two types of potentials, a cyclic potential and the Coulomb potential.
First, we studied behavior of a soliton in a cyclic potential.
Although our analysis does not predict detailed scattering of the soliton,
we find that this analysis gives insight to flux of the soliton,
and predicts the Laue pattern.
Second, we obtained a spectrum of eigenvalues in a system
in which a soliton is bounded on the Coulomb potential.
The eigenvalues are not interpreted as energy
since we did not quantize the system,
but we saw that the system absorbs external waves with
frequency described by the Rydberg formula.

Although we did not argue about spin in our analysis,
it is very interesting and important problem whether soliton models
can express spin or not.




\end{document}

%% file: oscillation.pstex_t
\begin{picture}(0,0)%
\includegraphics{oscillation.pstex}%
\end{picture}%
\setlength{\unitlength}{3947sp}%
\begingroup\makeatletter\ifx\SetFigFont\undefined%
\gdef\SetFigFont#1#2#3#4#5{%
  \reset@font\fontsize{#1}{#2pt}%
  \fontfamily{#3}\fontseries{#4}\fontshape{#5}%
  \selectfont}%
\fi\endgroup%
\begin{picture}(3963,1468)(289,-860)
\put(3976,-811){\makebox(0,0)[lb]{\smash{{\SetFigFont{12}{14.4}{\rmdefault}{\mddefault}{\updefault}{\color[rgb]{0,0,0}$x$}%
}}}}
\put(2701,-61){\makebox(0,0)[lb]{\smash{{\SetFigFont{12}{14.4}{\familydefault}{\mddefault}{\updefault}{\color[rgb]{0,0,0}wave length $\sim \infty$}%
}}}}
\put(1276, 89){\makebox(0,0)[lb]{\smash{{\SetFigFont{12}{14.4}{\familydefault}{\mddefault}{\updefault}{\color[rgb]{0,0,0}$\omega _0$}%
}}}}
\put(2176,464){\makebox(0,0)[lb]{\smash{{\SetFigFont{12}{14.4}{\rmdefault}{\mddefault}{\updefault}{\color[rgb]{0,0,0}density, energy or etc.}%
}}}}
\end{picture}%

%% file: boost.pstex_t
\begin{picture}(0,0)%
\includegraphics{boost.pstex}%
\end{picture}%
\setlength{\unitlength}{3947sp}%
\begingroup\makeatletter\ifx\SetFigFont\undefined%
\gdef\SetFigFont#1#2#3#4#5{%
  \reset@font\fontsize{#1}{#2pt}%
  \fontfamily{#3}\fontseries{#4}\fontshape{#5}%
  \selectfont}%
\fi\endgroup%
\begin{picture}(5463,2019)(289,-1273)
\put(2326,-211){\makebox(0,0)[lb]{\smash{{\SetFigFont{12}{14.4}{\familydefault}{\mddefault}{\updefault}{\color[rgb]{0,0,0}Lorentz boost}%
}}}}
\put(2176,-1111){\makebox(0,0)[lb]{\smash{{\SetFigFont{12}{14.4}{\familydefault}{\mddefault}{\updefault}{\color[rgb]{0,0,0}$x$}%
}}}}
\put(676,539){\makebox(0,0)[lb]{\smash{{\SetFigFont{12}{14.4}{\familydefault}{\mddefault}{\updefault}{\color[rgb]{0,0,0}$t$}%
}}}}
\put(4351,614){\makebox(0,0)[lb]{\smash{{\SetFigFont{12}{14.4}{\familydefault}{\mddefault}{\updefault}{\color[rgb]{0,0,0}$t$}%
}}}}
\put(5476,-436){\makebox(0,0)[lb]{\smash{{\SetFigFont{12}{14.4}{\familydefault}{\mddefault}{\updefault}{\color[rgb]{0,0,0}$x$}%
}}}}
\put(751,-736){\makebox(0,0)[lb]{\smash{{\SetFigFont{12}{14.4}{\familydefault}{\mddefault}{\updefault}{\color[rgb]{0,0,0}equivalent phase line}%
}}}}
\end{picture}%

%% file: potential.pstex_t
\begin{picture}(0,0)%
\includegraphics{potential.pstex}%
\end{picture}%
\setlength{\unitlength}{3947sp}%
\begingroup\makeatletter\ifx\SetFigFont\undefined%
\gdef\SetFigFont#1#2#3#4#5{%
  \reset@font\fontsize{#1}{#2pt}%
  \fontfamily{#3}\fontseries{#4}\fontshape{#5}%
  \selectfont}%
\fi\endgroup%
\begin{picture}(4824,2568)(589,-1873)
\put(1051,539){\makebox(0,0)[lb]{\smash{{\SetFigFont{14}{16.8}{\rmdefault}{\mddefault}{\updefault}{\color[rgb]{0,0,0}potential}%
}}}}
\put(976,-211){\makebox(0,0)[lb]{\smash{{\SetFigFont{14}{16.8}{\rmdefault}{\mddefault}{\updefault}{\color[rgb]{0,0,0}$0$}%
}}}}
\put(4051,-661){\makebox(0,0)[lb]{\smash{{\SetFigFont{14}{16.8}{\rmdefault}{\mddefault}{\updefault}{\color[rgb]{0,0,0}$p^{\prime}$}%
}}}}
\put(1801,-211){\makebox(0,0)[lb]{\smash{{\SetFigFont{14}{16.8}{\rmdefault}{\mddefault}{\updefault}{\color[rgb]{0,0,0}$p$}%
}}}}
\put(976,-1561){\makebox(0,0)[lb]{\smash{{\SetFigFont{14}{16.8}{\rmdefault}{\mddefault}{\updefault}{\color[rgb]{0,0,0}$-V_0$}%
}}}}
\end{picture}%

%% file: cyclic.pstex_t
\begin{picture}(0,0)%
\includegraphics{cyclic.pstex}%
\end{picture}%
\setlength{\unitlength}{3947sp}%
\begingroup\makeatletter\ifx\SetFigFont\undefined%
\gdef\SetFigFont#1#2#3#4#5{%
  \reset@font\fontsize{#1}{#2pt}%
  \fontfamily{#3}\fontseries{#4}\fontshape{#5}%
  \selectfont}%
\fi\endgroup%
\begin{picture}(4748,2316)(139,-1390)
\put(1726,-1336){\makebox(0,0)[lb]{\smash{{\SetFigFont{12}{14.4}{\familydefault}{\mddefault}{\updefault}{\color[rgb]{0,0,0}cyclic potential}%
}}}}
\put(151,389){\makebox(0,0)[lb]{\smash{{\SetFigFont{12}{14.4}{\familydefault}{\mddefault}{\updefault}{\color[rgb]{0,0,0}input soliton}%
}}}}
\put(2176,-511){\makebox(0,0)[lb]{\smash{{\SetFigFont{12}{14.4}{\familydefault}{\mddefault}{\updefault}{\color[rgb]{0,0,0}$\theta $}%
}}}}
\put(451,-1186){\makebox(0,0)[lb]{\smash{{\SetFigFont{12}{14.4}{\familydefault}{\mddefault}{\updefault}{\color[rgb]{0,0,0}reflection}%
}}}}
\put(3826,389){\makebox(0,0)[lb]{\smash{{\SetFigFont{12}{14.4}{\familydefault}{\mddefault}{\updefault}{\color[rgb]{0,0,0}output soliton}%
}}}}
\end{picture}%